\documentclass[12pt]{report}

\usepackage{amsmath}
\usepackage{graphicx}
\usepackage{geometry}
\usepackage{setspace}
\usepackage{fancyhdr}
\usepackage{graphicx}
\usepackage{caption}
\usepackage{subcaption}
\usepackage{siunitx}

\begin{document}
	
	\begin{titlepage}
		\begin{center}
			\vspace*{1in}
			
			\Huge
			\textbf{Photon Dynamics and Collision Risks in Relativistic Spaceflight: A Comparative Study of Methods and Implications}
			
			\vspace{0.5in}
			\Large
			by \\
			\vspace{0.2in}
			\textbf{Kevin Li}
			
			\vspace{0.5in}
			
			\Large
			\textbf{Supervisor:} \\
			\textbf{Dr. Joao Rodrigues}
			
			\vfill
			
			\Large
			A Dissertation Presented for the Degree of\\
			Bachelor of Science (B.Sc.)\\
			in\\
			Physics\\
			
			\vspace{0.5in}
			
			\Large
			The University of Hong Kong \\
			Faculty of Science \\
			16. Aug. 2024
			
		\end{center}
	\end{titlepage}
	
	\pagenumbering{roman}
	\section*{Acknowledgments}
	
	I would like to express my deepest gratitude to my supervisor, Dr. Joao Rodrigues, for his invaluable guidance, insightful feedback, and unwavering support throughout the course of this dissertation. His expertise and encouragement have been instrumental in shaping this work, and I am truly grateful for his mentorship. I also appreciate his patience and dedication in helping me navigate the challenges of this project.
	
	\newpage
	\section*{Introduction}
	
	Relativistic spaceflight, travelling at velocities approaching the speed of light, has been attractive to scientists and futurists alike since it poses both a possibility for interstellar travel as well as a massive technical challenge. When the spacecraft has reached a relatively high speed, the physical laws governing its motions diverge significantly from the Newtonian mechanics, which can be perfectly applied at a lower speed context. This dissertation discussed the challenges associated with relativistic spaceflight, focusing on the collision with other interstellar particles and the interactions with the cosmic microwave background radiation.
	
	\section*{Collision Dynamics in Relativistic Spaceflight}
	
	In the field of relativistic spaceflight, where a spacecraft's velocity approaches the speed of light, collisions with interstellar particles pose a significant challenge. Even the smallest particles can create extremely large extra kinetic energy at such high speeds, potentially leading to massive damage. There are two types of collisions I would like to illustrate:
	\begin{equation}
		\text{Two Types of Collisions:}
		\left\{
		\begin{array}{l}
			\text{Collisions with dust and other space objects.}\notag\\
			\text{Collisions with molecules.}\notag
		\end{array}
		\right.
	\end{equation}
	
	To understand this effect more intuitively, suppose the spacecraft's mass is $M$, the mass of the particle it collides with is $m$, and the spacecraft reaches velocity as $v$. According to the conservation of energy and momentum:
		\begin{equation}
			\left\{
			\begin{array}{l}
				E_{tot} = M\gamma c^2 + mc^2\notag\\
				p_{tot} = Mv\gamma\notag
			\end{array}
			\right.
		\end{equation}
	In order to minimize the damage, we would like to reach the state where, in the center of the mass frame, these two objects stop moving, leading to the equation:\[
	v'_{spacecraft} = v_{object} = v'
	\]
	then
	\begin{align}
		Mv\gamma &= (m+M) v'\gamma'\notag\\
		\Rightarrow \quad v' = Mv[&\frac{(m+M)^2}{\gamma^2}+M^2\cdot(1-\frac{1}{\gamma^2})]^{-\frac{1}{2}}\notag
	\end{align}
	Besides, the energy converted would be:
	\begin{align}
		\Delta E &= (\gamma-\gamma')Mc^2\notag\\
		&=(\frac{1}{m+M}\cdot\sqrt{m^2\cdot(1-\beta^2)+2mM(1-\beta^2)+M^2}-1)\cdot\gamma Mc^2\notag
	\end{align}
	However, this formula isn't very straightforward. Therefore, I would like to use the Taylor's expansion under the assumption ($\epsilon = \frac{m}{M}\rightarrow 0$). Hence, the formula can be simplified as\[
	\Delta E\approx [(\frac{m}{M})^2\cdot(\frac{5}{2}-\frac{3}{2}\cdot\beta^2)-\beta^2\frac{m}{M}]\cdot\gamma Mc^2
	\]
	According to the formula, the energy transferred to the spacecraft would be extremely large even if the mass $m$ is small, as shown in Figure I.
	
	Considering the smaller molecular level, the spaceship is flying through a region filled with hydrogen molecules. Although, in reality, it must be a mixture of multiple species of atomic molecules, for the sake of simplicity of illustration, let us assume that there are only hydrogen molecules in this region. Assuming that the density of hydrogen molecules in this region is n and the cross-sectional area of the spacecraft in this region in a plane perpendicular to its velocity is $A$, transferred energy could be expressed as
	\[
	\dot{E}\approx \oint_A (n_{H_2}\cdot v(\gamma-1)m_{H_2}\cdot c^2) dA
	\]
	
	The risk associated with these encounters is a matter of kinetic energy and the probability distribution over large interstellar distances. The methods for detecting, avoiding and protecting the spacecraft from being damaged are critical considerations for the design of relativistic spacecraft.
	
	\section*{Photon Interactions and the Schwinger Limit}
	As a spacecraft approaching a relatively high speed close to the speed of light, we have to consider the interaction between photons and nucleons. One of the most critical interactions I need to emphasise is a quantum mechanical process where a photon with sufficient energy interacts with a photon and creates an electron-positron pair. This phenomenon is of particular concern in relativistic spaceflight since it has set up speed restrictions on how fast a spacecraft can travel without experiencing significant damage.
	
	If we simply consider the conservation of energy, we could derive the equations as follow:
	\begin{align}
		h\nu&\ge2m_e\cdot c^2\notag\\
		\nu&\ge\frac{2m_e\cdot c^2}{h}\approx 2.47\times 10^{20}\  \si{Hz}.\notag
	\end{align}
	$m_e$ is the mass of an electron.
	
	This equation could only be objective in $S'$ frame. We need to transfer it into the rest frame,
	\begin{align}
		\sqrt{\frac{1+\beta}{1-\beta}}\cdot 10^{12} &= \nu\notag\\
		v &\approx (1-3.3\times 10^{-17})\cdot c\notag
	\end{align}
	There is one thing I shall clarify: the term $10^{12}$ is given by approximately the average frequency of the (Planck) distribution of CMB photons, which is $160G\si{Hz}$ (Ulvi Yurtsever et al., 2015). As the spacecraft moves at relativistic speeds, the CMB photons are blue-shifted in the spacecraft's frame, and the energy is increased. Fortunately, this blue-shifting effect significantly decreases the required velocity, as the energy of the photons quickly exceeds the boundary energy required for pair production.
	
	In addition, the Schwinger limit is another fundamental challenge in relativistic spaceflight. This phenomenon occurs when the energy density of the electromagnetic field exceeds a specific limit, leading to the vacuum breakdown into electron-positron pairs. Here, I am going to offer two kinds of approaches.
	
	First, the energy density is necessary for my derivation. From Maxwell's equations(two of those which I will express in Gaussian units):
	\begin{equation}
		\left\{
		\begin{array}{l}
			\vec{\nabla}\cdot\vec{E} = 4\pi\rho\notag\\
			\vec{\nabla}\times\vec{B} = \frac{1}{c}\cdot\frac{\partial}{\partial t}\vec{E}+\frac{4\pi}{c}\cdot\vec{J}
		\end{array}
		\right.
	\end{equation}
	To derive the energy density, start from $W$, which is the electron potential energy.
	\begin{align}
		W &= \int \rho(\vec{r})\cdot U(\vec{r}) d^3r \notag\\
		&= \int \frac{1}{4\pi}(\vec{\nabla}\vec{E})\cdot U(\vec{r}) d^3r\notag\\
		&= \int \frac{1}{8\pi}E^2 d^3r\notag
	\end{align}
	The energy density is considered by taking a square with side length $\delta$ on the surface of the spaceship, sacrificing the relationship:
	\[
	\delta^3\cdot u = h\nu
	\]
	Therefore, the energy density will be extracted from the equation above: $u = \frac{E^2}{8\pi} = \frac{h\nu}{\delta^3}$.
	
	From the definition of the Schwinger limit, it corresponds to the electric field strength at which spontaneous electron-positron pair production from the electromagnetic field(photon). Then
	\begin{align}
		e\cdot E_S\frac{h}{m_e\cdot c} &= 2m_ec^2\notag\\
		\Rightarrow E_S = \frac{m_e^2c^3}{\pi e\hbar}&\approx 1.4\times 10^{14}\ \si{(erg/cm^3)^{1/2}}\notag
	\end{align}
	Here, $h/m_ec$ is the electron Compton wavelength (Ulvi Yurtsever et al., 2015). The phenomenon will occur when the electric field $E\le E_S$, giving us
	\begin{align}
		E\approx\sqrt{\frac{8\pi h\nu}{\delta^3}}&\ge E_S = \frac{m_e^2\cdot c^3}{\pi e\hbar}\notag\\
		\Rightarrow\nu\ge\frac{m_e^4c^6\delta^3}{16\pi^2e^2\hbar^3} &= \frac{E_S^2}{16\pi^2\hbar}\delta^3\approx7.8\times10^{51}\ \si{Hz\times[\delta(cm)]^3}\notag
	\end{align}
	
	Here is another method to approach it: rather than considering the cubic volume, let's focus on the energy density. Denote the angle between the velocity of the spacecraft and the momentum of the background photons is $\theta$. Suppose the energy of a photon is $E$ in the rest frame and $E'$ in the spacecraft's moving frame, which corresponds to each other by
	\[
	E = E'\cdot\gamma\cdot(1+\beta\cos\theta)
	\]
	Since the requirement is $E\ge2m_ec^2$, we could obtain the energy in $S'$ frame,\[
	E'\ge\frac{2m_ec^2}{\gamma\cdot(1+\beta\cdot\cos\theta)}
	\]
	We have already derived the energy density, which is $u_S = \frac{E^2_S}{8\pi}$.
	To obtain the Schwinger's limit, we need $u_{eff}\ge u_S$, and $u_{eff}$ is actually the energy density in the $S'$ frame.\[
	u_{eff} = \rho\cdot E'
	\]
	$\rho$ is the number density of CMB photons. Then
	\begin{align}
		\rho\cdot E\cdot(1+\beta\cos\theta)\cdot\gamma&\ge\frac{E_S^2}{8\pi}\notag\\
		E = h\nu&\ge\frac{E_S^2}{8\pi\rho\gamma\cdot(1+\beta\cos\theta)}\notag
	\end{align}
	These approaches complement each other, with the first providing a more traditional, field-based perspective while the second offers insights into how relativistic dynamics modify these conditions.
	
	\section*{Advanced Photon Interactions in Relativistic Spaceflight}
	 Photon interactions are a common and fundamental challenge in the field of relativistic spaceflight, where the speed of the spacecraft approaches the speed of light. Though we have focused on discussing pair production before, it's not the only critical challenge we need to overcome. Other photon interaction mechanisms, such as Compton scattering and photon absorption, also require careful consideration. These interactions can have significant implications for both thermal management and material designing, which need more advanced engineering solutions.
	 
	 Compton scattering is a process in which photons transfer part of the energy to electrons. This process will become more and more critical as the velocity of the spacecraft approaches relativistic speed. The frequency of the photons has been changed due to the Doppler effect, leading to the increased energy of photons in the spacecraft's frame. This frequency shift means that even if the photons originally had lower energy in the rest frame, they could gain enough energy to cause substantial Compton scattering, leading to several potential challenges.
	 
	 The first challenge is the thermal buildup. Suppose we denote the scattering angle as $\theta$, $E$ is the initial photon energy, and $m_e$ is the mass of the electron at rest frame. Therefore, the energy transferred from photons to electrons results in heating the spacecraft's surface. Consider a photon with initial energy $E$ and momentum $p = \frac{E}{c}$ experiencing the collision with a stationary electron. The total energy before and after the collision must be conserved:\[
	 E + m_ec^2 = E' + E_e
	 \]
	 $E'$ and $E_e$ are the total energies of the photon and the electron after the collision respectively. Using the relativistic energy-momentum relation:\[
	 E_e = \sqrt{(p_e\cdot c)^2 + (m_e\cdot c^2)^2}
	 \]
	 with the conservation of momentum for the system in the $x$ and $y$ directions,
	 \begin{align}
	 	\text{$x$-direction:}\ &\frac{E}{c} = \frac{E'}{c}\cdot\cos\theta + p_e\cdot\cos\phi\notag\\
	 	\text{$y$-direction:}\ &0 = \frac{E'}{c}\cdot\sin\theta - p_e\sin\phi\notag
	 \end{align}
	 where $\phi$ is the angle of the recoiling electron relative to the initial photon direction, we could derive the relationship between the $\lambda'$ and the $\lambda$, which are the initial and final wavelengths of the photons respectively, together:
	 \begin{equation}
	 	\left\{
	 	\begin{array}{l}
	 		\frac{E^2}{c^2} = \frac{E'^2}{c^2} + p_e^2 + 2\cdot\frac{E'}{c}p_e\cos\theta\notag\\
	 		p_e^2c^2 = (E + m_e c^2 - E')^2 - m_e^2c^4\notag
	 	\end{array}
	 	\right.
	 \end{equation}
	 
	 By substituting $p_e^2$ and simplifying the equation, we arrive the final result: 
	 \[
	 \lambda' -\lambda = \frac{h}{m_ec}\cdot(1-\cos\theta)
	 \]
	 
	 Since we can always substitute the wavelength $\lambda$ with $\frac{hc}{E}$. The energy of the scattering photons $E'$ can be described by the Compton equation:\[
	 E' = \frac{E}{1+\frac{E}{m_ec^2}\cdot(1-\cos\theta)}
	 \]
	 
	 The energy of the incoming photons will be increased due to the relativistic effects, the resulting thermal energy can be significant. Hence, advanced cooling mechanics are most likely to be needed.
	 
	 Another factor we should consider is the momentum transfer. It's easy to deduce that the momentum transfer $\Delta p$ to the electron, and thus to the spacecraft, during Compton scattering can be expressed as:\[
	 \Delta p = \frac{E}{c}\cdot(1-\cos\theta)
	 \]
	 
	 One thing should be noted: this is the contribution of a single photon. Although it appears very small for a spacecraft, after we have accumulated the influence, the effect will never be neglected anymore. To approach to this, we shall need the help of the Planck's distribution of photon energies:
	 \[
	 n(\nu)d\nu = \frac{8\pi\nu^2}{c^3}\cdot\frac{1}{\exp(h\nu/k_B\cdot T)-1}d\nu
	 \]
	 Therefore, the integrated total momentum transfer $P_{total}$ is found by integrating over all incident photons across the spacecraft's cross-sectional area $A$, which is perpendicular to the spacecraft's velocity, and over the range of frequencies in the Planck distribution:\[
	 P_{total} = \int\!\int_A (\int_{0}^{\infty}\Delta p(\nu)\cdot n(\nu) \cdot d\nu)
	 \]
	 
	 Let's simplify this expression:
	 \begin{align}
	 	P_{total} &= \frac{8\pi hA}{c^4}(\int_{0}^{\infty}\frac{\nu^3}{\exp(h\nu/k_BT)-1}d\nu)\cdot(1-\cos\theta)\notag\\
	 	&= \frac{8\pi hA}{c^4}\cdot[\frac{(k_BT)^4}{h^3}\cdot\frac{\pi^4}{15}]\cdot(1-\cos\theta)\notag\\
	 	&= \frac{8\pi^5k_B^4T^4A}{15c^4h^3}\cdot(1-\cos\theta)\notag
	 \end{align}
	 The reason why we can remove the double integration is the distribution and the transfer are uniformly distributed on the surface. This expression represents the integrated momentum transfer from all CMB photons impinging on the spacecraft.
	 
	 Apart from that, another key interaction to consider is photon absorption, where incoming photons are absorbed by the spacecraft's material, leading to the increase of the temperature. Since this absorbed energy perhaps will be emitted later in the form of radiation, it might lead to new challenges.
	
	\section*{Cosmic Microwave Background (CMB) Photon Distribution and Relativistic Effects}
	The cosmic microwave background radiation provides a pervasive backdrop to any object moving through space, including the relativistic spacecraft. Under this framework, this radiation appears to have a nearly uniform isotropic distribution of photons, and the temperature is approximately $2.73 \si{K}$. However, the momentum distribution of the photons will be severely changed due to the effects of Lorentz transformation.
	
	Using the Planck density, which can be expressed as\[
	n(\vec{k})d^3k = \frac{1}{4\pi^3}\frac{1}{e^{\hbar\omega/k_BT}-1}d^3k
	\]
	However, although this formula describes the distribution clearly, gaining the average number of photons per volume is still quite complicated due to the triple integration. Therefore, we use the frequency $\nu$ rather than the position vector $\vec{k}$.
	\begin{align}
		\rho &= \int_{0}^{\infty}n(\nu)d\nu \notag\\
		&= \frac{2k_B^3T^3}{\pi^2c^3\hbar^3}\cdot\zeta(3)\notag\\
		&\approx 412\ \text{photons/cm}^3\notag
	\end{align}
	This provides the distribution of photons in the rest frame. In order to transfer it into another frame, which is the spacecraft's frame, we still have to choose the formula $n(\vec{k})$ and consider the relationship between $\vec{k}$ and $\vec{k}'$, which is given by
	\begin{equation}
		\left\{
		\begin{array}{c}
			\hat{n}(\vec{k}') = n[\vec{k}(\vec{k}')]\cdot |\frac{\partial \vec{k}}{\partial \vec{k}'}|\notag\\
			\notag\\
			\vec{k}(\vec{k}') = [\gamma\cdot(k_x'+v\cdot\omega'), k_y', k_z']\notag\\
			\notag\\
			|\frac{\partial \vec{k}}{\partial\vec{k}'}| = \gamma(1+\frac{v\cdot k_x'}{\omega'})\notag
		\end{array}
		\right.
	\end{equation}
	Suppose the distribution of the planets in the universe could be regarded as a uniform function. In that case, as the spacecraft moves with relatively high velocity under the spherical coordinate, the radiation distribution of planets is irrelevant to the azimuth angle $\phi$. Therefore, we could derive a temperature function only with respect to polar angle $\theta$. From the Doppler's effect,
	\[
	\nu' = \nu\cdot\gamma\cdot(1-\beta\cdot\cos\theta)
	\]
	and\[
	T(\theta) = \frac{T}{\gamma\cdot(1-\beta\cos\theta)}
	\]
	In a fast-moving spacecraft, the spectral images we see from the radiation will be concentric circles. However, this image would be based on the fact that the stars are uniformly distributed throughout the universe. If we adjust the distribution function, the image will adjust as well. For instance, if we take the distribution of the function is $\rho(\theta,\phi) = 1 + Asin(2\theta)\cdot\cos(2\phi)+ Be^{-\theta/\theta_0}$, then the diagram would be seen as in the figure 2. The circle would be stretched and squeezed due to the aberration of light.
	
	Furthermore, we would like to pay attention to the spectral intensity of the reflected radiation in the CMB rest frame. There are two methods to derive it.
	
	Suppose we set up a coordinate system in the CMB rest frame. In that case, we denote that the radiation is detected by an observer at $\vec{r}$ at time $t$, the actual time when the radiation is reflected by the spacecraft at $\vec{r}_1 = \vec{v}\cdot\sigma$, in which $\sigma$ is the retarded time, and $A$ is an area element which would appear as a point source. Therefore, the equation related to the $\sigma$ is
	\[
	c(t-\sigma) = \sqrt{(x-v\sigma)^2+y^2+z^2}
	\]
	The solution will be (Ulvi Yurtsever et al., 2015)\[
	\sigma = \gamma^2(t-\frac{vx}{c^2}-\frac{1}{c\gamma}\sqrt{\gamma^2(x-vt)^2+y^2+z^2})
	\]
	To derive the formula of spectral intensity, there are several terms we shall take into consideration:
	\begin{enumerate}
		\item[] Geometric spreading factor:\[\frac{1}{4\pi|\vec{r}-\vec{v}\sigma|^2}\]
		
		\item[] 
		
		\item[] Frequency-dependent component:\[\frac{\omega^2}{\pi^2c^2}\]
		
		\item[] 
		
		\item[] Angular-dependent expression under the Lorentz's formula:
		\[\gamma^2\left[1+\left(\frac{\vec{v}}{c|\vec{r}-\vec{v}\sigma|}\cdot(\vec{r}-\vec{v}\sigma-2[(\vec{r}-\vec{v}\sigma)\cdot\hat{s}]\hat{s})\right)\right]^2\]
		
		\item[]
		
		\item[] Modified Planck distribution by relativistic effects:
		\[\exp\left[\frac{\gamma^2\hbar\omega}{k_BT}\left[1+\left(\frac{\vec{v}}{c|\vec{r}-\vec{v}\sigma|}\cdot\left(\vec{r}-\vec{v}\sigma-2[(\vec{r}-\vec{v}\sigma)\cdot\hat{s}]\cdot\hat{s}\right)\right)\right]^2\right]-1\]
	\end{enumerate}
	$\hat{s}$ is the unit vector which is the norm vector of the surface $A$.
	
	Since the intensity $I$ of a wave decreases in proportion to the square of the distance $r$ from the source:\[
	I(r) = \frac{I_0}{r^2}
	\]
	
	This relationship holds when the wave propagates spherically outward from a point source in a homogeneous, isotropic medium. In the context of radiation, such as electromagnetic radiation from a relativistic spacecraft, this geometric spreading factor accounts for how the energy of emitted photons is diluted as they spread out from the point of emission. Here, the geometric spreading factor is represented by:
	\[
	\frac{1}{4\pi|\vec{r}-\vec{v}\sigma|^2}
	\]
	in which the factor $|\vec{r}-\vec{v}\sigma|$ represents the distance between the observer and the source of radiation (reflected point), and $4\pi$ represents the total solid angle over which the radiation is spread in three-dimensional space.
	
	Combining these terms together, we have gained the final result of the spectral intensity on $A$ spot, which is\[
	I(\vec{r},t,\omega) = \frac{A}{4\pi|\vec{r}-\vec{v}\sigma|^2}\frac{\omega^2}{\pi^2c^2}\frac{\gamma^2[1+(\frac{\vec{v}}{c|\vec{r}-\vec{v}\sigma|}\cdot(\vec{r}-\vec{v}\sigma-2[(\vec{r}-\vec{v}\sigma)\cdot\hat{s}]\hat{s}))]^2}{\exp[\frac{\gamma^2\hbar\omega}{k_BT}[1+(\frac{\vec{v}}{c|\vec{r}-\vec{v}\sigma|}\cdot(\vec{r}-\vec{v}\sigma-2[(\vec{r}-\vec{v}\sigma)\cdot\hat{s}]\cdot\hat{s}))]^2]-1}
	\]
	this expression seems a little bit awful. Therefore, I would like to propose another approach to the result. Let's start with the Planck's Law for Spectral Radiance:\[
	I_{\omega}(T) = \frac{\omega^2}{\pi^2c^2}\cdot\frac{\hbar\omega}{\exp(\hbar\omega/k_BT)-1}
	\]
	After transferring it into the moving frame, using the fact that $\omega' = \omega\cdot\gamma\cdot(1-\beta\cdot\cos\theta)$, we could derive the formula as\[
	I_{\omega}'(T) = I_{\omega}(T)\cdot\left(\frac{\omega'}{\omega}\right)^3 = \frac{\omega^2}{\pi^2c^2}\cdot\frac{\hbar\omega}{\exp(\hbar\omega/k_BT)-1}\cdot\gamma^3\cdot(1-\beta\cdot\cos\theta)^3
	\]
	After taking the geometric spreading and relativistic effects into consideration, we obtain the final formula:\[
	I(\vec{r},\omega,t) = \frac{A}{4\pi|\vec{r}-\vec{v}\sigma|^2}\cdot\frac{\omega^2}{\pi^2c^2}\cdot\frac{\hbar\omega}{\exp(\hbar\omega/k_BT)-1}\cdot\gamma^3\cdot(1-\beta\cdot\cos\theta)^3
	\]
	
	This method is more straightforward to apply, focusing on the Doppler shift and relativistic scaling of intensity. It is a good choice when simplicity is desired or when the first method's additional complexities are unnecessary.
	
	Besides, as the spacecraft moves through the CMB, the frequency of the photons observed in the spacecraft's frame $\nu'$ is Doppler-shifted. We have derived the transfer relationships between the frequency $\nu$ and temperature $T$ in different frames as follow:
	\begin{align}
		\nu' = \nu\cdot\gamma\cdot(1-\beta\cos\theta)\notag\\
		T'(\theta) = \theta\cdot\gamma\cdot(1-\beta\cos\theta)\notag
	\end{align}
	After we have substitute it into the Planck's law, the spectral intensity in the rest frame is given by:\[
	I(\nu,T) = \frac{2h\nu^3}{c^2}\cdot\frac{1}{\exp(h\nu/k_BT)-1}
	\]
	Therefore, the black body spectrum will be:
	\begin{align}
		I'(\nu',T') &= \frac{2h\nu'^3}{c^2}\cdot\frac{1}{\exp(h\nu/k_B T(\theta))-1}\notag\\
		&= \frac{2h\cdot(\nu\cdot\gamma\cdot(1-\beta\cos\theta))^3}{c^2}\cdot\frac{1}{\exp\left[h\cdot\nu\cdot\gamma\cdot(1-\beta\cos\theta)/k_B\cdot T\cdot\gamma\cdot(1-\beta\cos\theta)\right]-1}\notag
	\end{align}
	
	There is another effect we shall consider: The sunyaev-Zeldovich effect, which is a phenomenon where CMB photons gain energy through inverse Compton scattering with electrons in galaxy clusters; therefore, the CMB spectrum would be distorted. The magnitude of the distortion in the Sunyaev-Zeldovich effect is closely related to the temperature of the intracluster gas and the frequency of scattering events within the cluster, as higher temperatures and more frequent scatterings increase the energy transfer from electrons to CMB photons, leading to a more pronounced shift in the observed CMB spectrum (Birkinshaw, 1999; Carlstrom et al., 1996).
	
	CMB photons are shifted to higher energies, and this is predictable by modelling the spacecraft's interactions with the CMB, which is an SZ-like effect. The difference in the CMB intensity caused by interactions such as those modelled in the SZ effect can be expressed by integrating the difference between the observed modified intensity and the original CMB intensity over the solid angle, as detailed by Sunyaev and Zeldovich (1970) and further explored in the context of galaxy clusters by Birkinshaw (1999).
	
	Their works give out the mathematical expression of the total shift in the energy of the CMB photons due to interaction with the spacecraft:\[
	\Delta I_{CMB}(\nu) = \int_\Omega[I'(\nu',T')-I(\nu,T_{CMB})]d\Omega
	\]
	
	The shift in CMB temperature and spectrum has direct implications for both and detection of the spacecraft and its thermal management:
	\begin{itemize}
		\item[] \textbf{Spacecraft detection:}
		
		The spectral distortion is caused by the spacecraft's motion, which we can observe on Earth in a similar way we do to detect the SZ effect with sensitive instruments. Therefore, this is a potential method for us to detect and track the relativistic spacecraft.
		\item[] \textbf{Thermal management:}\\
		Due to the Doppler effect, CMB photons will be blue-shifted, which is an increased thermal load on the spacecraft. Hence, there will be extra heat to be handled, which we can deal with using advanced thermal management. It's critical to adjust the direction of travel relative to the CMB.
	\end{itemize}
	
	Spacecraft designers can understand the risks and challenges brought by high-speed travel through the CMB by comparing the simulated CMB interactions with observational data from the SZ effect and developing strategies to minimize the potential risks.
	
	\section*{Design Considerations for Relativistic Spacecraft}
	Given all the challenges above, the design of relativistic spacecraft must contain advanced materials and systems to reduce the effect of the interactions of light. There are several approaches to consider:
	\begin{itemize}
		\item[] \textbf{High-reflectivity materials:}\\
		Spacecraft should be constructed using materials with high reflectivity across a broad spectrum of photon energies to minimize both absorption and Compton scattering.
		
		\item[] \textbf{Active Thermal Management:}\\
		It's critical to have effective thermal management systems to deal with the heat produced by the interactions between photons. These systems might include radiative heat sinks, liquid cooling circuits, and heat pipes that can redistribute excess heat to less critical areas of the spacecraft.
		
		\item[] \textbf{Photon Shields and Deflectors:}\\
		Detectors and deflectors on the spacecraft are essential to redirect incoming photons away from sensitive areas.
	\end{itemize}
	
	\section*{Conclusion}
	The exploration of relativistic spaceflight reveals various challenges, both theoretical and practical. Throughout this dissertation, we have delved into several critical aspects of these challenges, focusing on the collision between the spacecraft and objects, particles and photons, as well as the impact of cosmic microwave background radiation on a spaceship travelling at a speed which is close to the speed of the light. We also compared two methods for deriving different values and provided an analysis of these, offering the choice of method depending on the specific requirements of the analysis.
	
	In conclusion, this dissertation has provided a comprehensive examination of the challenges of relativistic spaceflight, offering new perspectives on resolving the problems we might encounter.
	\newpage
	\listoffigures
	\newpage
	\begin{figure}[h!]
		\centering
		\includegraphics[width=0.8\textwidth]{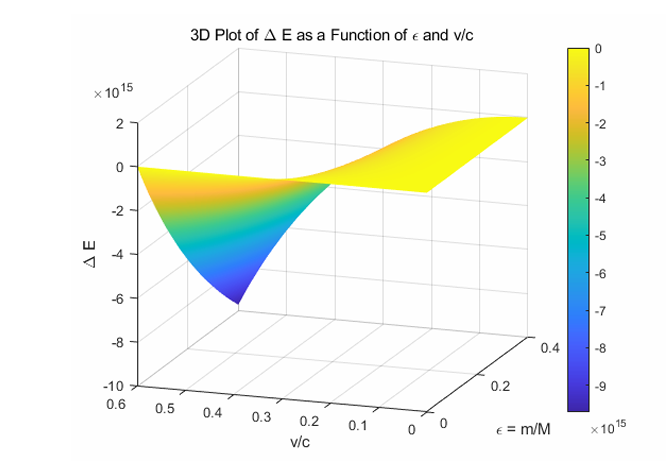}
		\caption{3D Plot of \(\Delta E\) as a Function of \(\epsilon = \frac{m}{M}\) and \(\beta = \frac{v}{c}\)}
		\label{fig:DeltaEPlot}
	\end{figure}
	\begin{figure}[h!]
		\centering
		\includegraphics[width=0.8\textwidth]{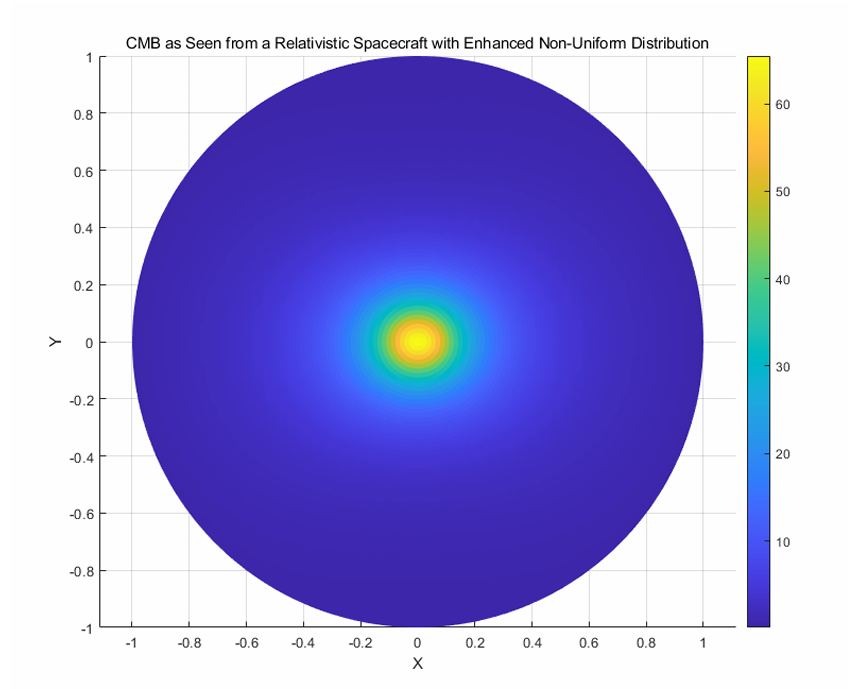}
		\caption{CMB as Seen from a Relativistic Spacecraft with Non-Uniform Distribution.}
		\label{fig:CMB}
	\end{figure}
	\newpage
	
	\pagenumbering{arabic}
	\setcounter{page}{1}
	
	\nocite{*}
	
	\addcontentsline{toc}{chapter}{References}
	\bibliographystyle{plain}
	\bibliography{Reference}
	
	
\end{document}